\documentclass[12pt]{article}
\usepackage{rotating}
\usepackage{longtable}
\usepackage{amsmath}
\usepackage{graphicx}
\usepackage{dcolumn}  
\usepackage{subfigure}
\usepackage{multirow}

\graphicspath{{ps}}

\newcommand{\mev}{\,\mathrm{MeV}}

\newcommand{\mevm}{\mathrm{MeV}/c^2}

\newcommand{\gevm}{\mathrm{GeV}/c^2}

\newcommand{\ee}{e^+e^-}
\newcommand{\uu}{\mu^+\mu^-}
\newcommand{\pp}{\pi^+\pi^-}
\newcommand{\U}{\Upsilon}
\newcommand{\Uf}{\Upsilon(5S)}
\newcommand{\Uo}{\Upsilon(1S)}
\newcommand{\Un}{\Upsilon(nS)}
\newcommand{\Ut}{\Upsilon(2S)}
\newcommand{\Uth}{\Upsilon(3S)}

\newcommand{\mmpp}{MM(\pi^+\pi^-)}
\newcommand{\mmp}{MM(\pi)}

\newcommand{\hb}{h_b(1P)}
\newcommand{\hcone}{h_c(1P)}
\newcommand{\hbp}{h_b(2P)}
\newcommand{\hbn}{h_b(mP)}

\newcommand{\ks}{K^0_S}

\newcommand{\pip}{\pi^{+}}
\newcommand{\pipm}{\pi^{\pm}}
\newcommand{\pim}{\pi^{-}}

\newcommand{\fb}{\mathrm{fb}^{-1}}

\newcommand{\etal}{\em et al.}

\newcommand{\dmhf}{\Delta M_{\rm HF}}

\newcommand{\zb}{Z_b}
\newcommand{\zbo}{Z_b(10610)}
\newcommand{\zbt}{Z_b(10650)}

\newcommand{\mzahb}{10605.1\pm2.2\,^{+3.0}_{-1.0}}
\newcommand{\gzahb}{11.4\,^{+4.5}_{-3.9}\,^{+2.1}_{-1.2}}
\newcommand{\mzbhb}{10654.5\pm2.5\,^{+1.0}_{-1.9}}
\newcommand{\gzbhb}{20.9\,^{+5.4}_{-4.7}\,^{+2.1}_{-5.7}}
\newcommand{\ahb}{1.8\,^{+1.0}_{-0.7}\,^{+0.1}_{-0.5}}
\newcommand{\phihb}{188\,^{+44}_{-58}\,^{+4}_{-9}}

\newcommand{\mzahbp}{10596\pm7\,^{+5}_{-2}}
\newcommand{\gzahbp}{16\,^{+16}_{-10}\,^{+13}_{-4}}
\newcommand{\mzbhbp}{10651\pm4\pm2}
\newcommand{\gzbhbp}{12\,^{+11}_{-9}\,^{+8}_{-2}}
\newcommand{\ahbp}{1.3\,^{+3.1}_{-1.1}\,^{+0.4}_{-0.7}}
\newcommand{\phihbp}{255\,^{+56}_{-72}\,^{+12}_{-183}}

\newcommand{\jpsi}{\mbox{$ J/\psi$}}

\newcommand\pubdate{\today}

\newcommand\pubnumber{}

\def\Title#1{\begin{center} {\Large #1 } \end{center}}
\def\Author#1{\begin{center}{ \sc #1} \end{center}}
\def\Address#1{\begin{center}{ \it #1} \end{center}}

\newcommand\pubblock{\rightline{\begin{tabular}{l} \pubnumber\\
         \pubdate  \end{tabular}}}
\newenvironment{Abstract}{\begin{center}{\bf Abstract}\end{center} \bigskip \begin{quotation}  }{\end{quotation}}
\newenvironment{Presented}{\begin{quotation} \begin{center}
             PRESENTED AT\end{center}\bigskip
      \begin{center}\begin{large}}{\end{large}\end{center} \end{quotation}}

\def\beq{\begin{equation}}
\def\eeq#1{\label{#1}\end{equation}}
\def\eeqn{\end{equation}}

\def\beqa{\begin{eqnarray}}
\def\eeqa#1{\label{#1}\end{eqnarray}}
\def\eeqan{\end{eqnarray}}

\let\bar=\overbar

\def\etal{{\it et al.}}

\def\Dslash{\not{\hbox{\kern-4pt $D$}}}
\def\dslash{\not{\hbox{\kern-2pt $\del$}}}

\def\ee{e^+e^-}

\def\msb{{\bar{\ssstyle M \kern -1pt S}}}

\begin{document}
\begin{titlepage}
\pubblock

\vfill


\Title{Observation of the $h_b$ states and Beyond}
\vfill \Author{A.E. Bondar} \Address{Budker Institute of Nuclear
Physics, Novosibirsk, 630090, Russia\\ On behalf of Belle
Collaboration} \vfill


\begin{Abstract}

Originally designed for CP violation studies in the B meson system,
the B-Factories recently showed an exciting capability for improving
our experimental knowledge in the field of hadron spectroscopy. We
review results on bottomonium spectroscopy from the Belle experiment
at the KEK-B $e^+e^-$ collider and present exciting new results from the
unique large data set taken at the $\Upsilon(5S)$ resonance.

\end{Abstract}

\vfill

\begin{Presented}
The Ninth International Conference on\\
Flavor Physics and CP Violation\\
(FPCP 2011)\\
Maale Hachamisha, Israel,  May 23--27, 2011
\end{Presented}
\vfill

\end{titlepage}
\def\thefootnote{\fnsymbol{footnote}}
\setcounter{footnote}{0}
%


\section{Introduction}

The Belle Collaboration has collected a large sample of $\ee$
collisions at the energy of the $\Uf$ resonance, which lies above
the threshold for production of $B_s$ meson pairs, primarily with a
purpose of studying decays of $B_s$. There have been a number of
unexpected results on the non-$B_s\bar{B}_s$ decays of the $\Uf$. In
particular, anomalously large rates for dipion transitions to lower
bottomonium states $\Uf\to(\Uo,\Ut,\Uth)\pp$ have been
observed~\cite{5s_rate}. If these signals are attributed entirely to
the $\Uf$ decays, the measured partial decay widths
$\Gamma[\Uf\to\U(nS)\pp]\sim0.5\,\mev$ are about two orders of
magnitude larger than typical widths for dipion transitions among
$\U(nS)$ states with $n\leq4$.
%

Recently the CLEO-c Collaboration observed the process
$\ee\to\hcone\pp$ at a rate comparable to the process
$\ee\to\jpsi\pp$ at $\sqrt{s}=4170\,\mev$ and found an indication of
an even higher transition rate at the $Y(4260)$
energy~\cite{Mitchell}. This implies that the $\hbn$ production
might be enhanced in the region of the $Y_b$ and motivates a search
for the $\hbn$ in the $\Uf$ data.

We use the full $\Uf$ data sample with the integrated luminosity of
$121.4\,\fb$ collected near the peak of the $\Uf$ resonance with the
Belle detector~\cite{BELLE_DETECTOR} at the KEKB asymmetric-energy
$\ee$ collider~\cite{KEKB}.


\section{Observation of the \boldmath $h_b(1P)$ and $h_b(2P)$}

\begin{figure*}[bth]
\includegraphics[width=5cm]{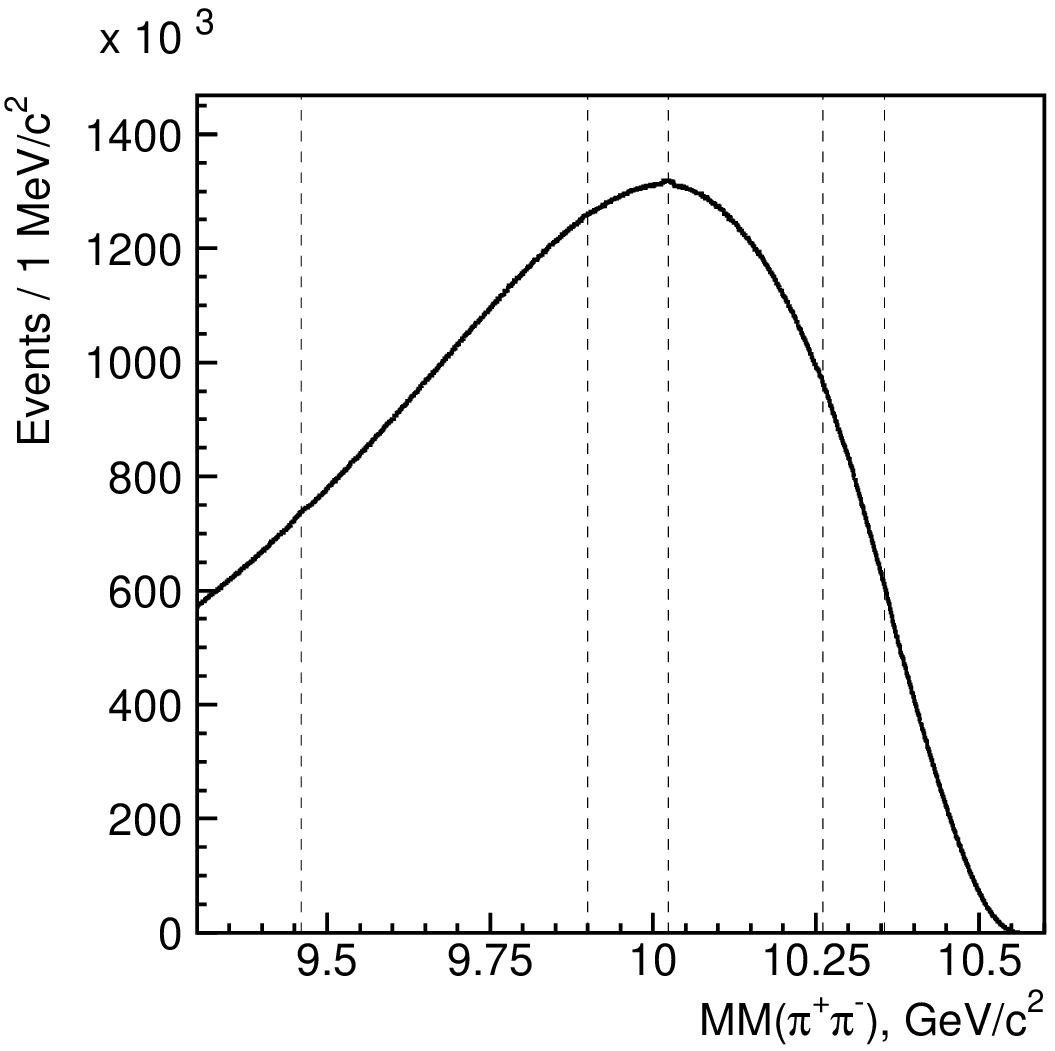}\includegraphics[width=10cm]{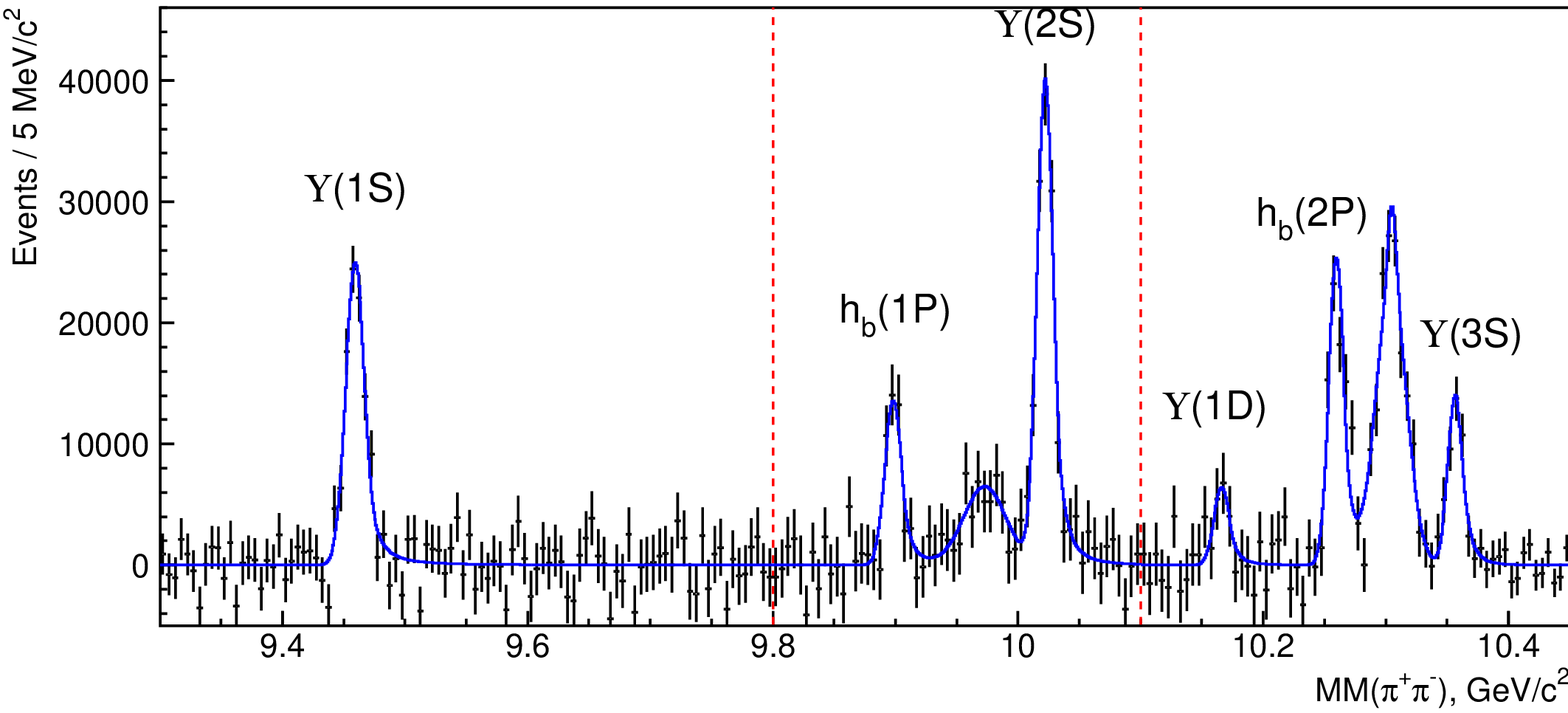}
\caption{(a)The $\mmpp$ distribution for the selected $\pp$
  pairs. Vertical lines indicate the locations of the $\Uo$, $\hb$,
  $\Ut$, $\hbp$ and $\Uth$ signals. (b) The $\mmpp$ spectrum with the combinatorial background and
  $\ks$ contributions subtracted (dots with error bars) and signal
  component of the fit function (solid histogram). The vertical dashed
  lines indicate the boundaries of the fit regions.}
\label{mmpp_all}
\end{figure*}

We observe the $\hb$ and $\hbp$ in the missing mass spectrum of
$\pp$ pairs. The $\pp$ missing mass is defined as $ MM(\pp) \equiv
\sqrt{(E_{c.m.}-E_{\pp}^*)^2-p_{\pp}^{*2}}, $ where $E_{c.m.}$ is
the center-of-mass (c.m.) energy, $E_{\pp}^*$ and $p_{\pp}^*$ are
the $\pp$ energy and momentum measured in c.m. frame. The details of
the analysis can be found in~\cite{mizprl}. The $MM(\pp)$
distribution for the selected $\pp$ pairs is shown in
Fig.~\ref{mmpp_all}(a). In this figure only the $\Uf\to\Uo\pp$ and
$\Uf\to\Ut\pp$ transitions are discernible.

To fit the $\mmpp$ spectrum, we separate it into three adjacent
regions with boundaries at $\mmpp=9.3\,\gevm$, $9.8\,\gevm$,
$10.1\,\gevm$ and $10.45\,\gevm$. We fit every region separately to
better control the complicated shape of the combinatorial
background, which is described by a Chebyshev polynomial of 6-7th
order. In region 3 we subtract the $\ks$ contribution bin-by-bin,
while in other regions its shape is smooth and is absorbed into
combinatorial background. The signal peaks are described by
Gaussians with paramenters obtained from  exclusive decays of the
$\U(nS)$ to $\uu$.
 The $\mmpp$
spectrum with the combinatorial background and $\ks$ contributions
subtracted, and the signal function resulting from the fit overlaid,
are shown in Fig.~\ref{mmpp_all}(b).
The significance of the $\hb$ and $\hbp$ signals which includes the
systematic uncertainty is $5.5\sigma$ and $11.2\sigma$,
respectively.

This is the first observation of the $\hb$ and $\hbp$ spin-singlet
bottomonium states in the reaction $\ee\to\hbn\pp$ at the $\Uf$
energy. We measure the masses and the cross sections relative to the
$\ee\to\Ut\pp$ cross-section:
$M=9898.25\pm1.06^{+1.03}_{-1.07}\,\mevm$,
$R=0.407\pm0.079^{+0.043}_{-0.076}$ for the $\hb$ and
$M=10259.76\pm0.64^{+1.43}_{-1.03}\,\mevm$,
$R=0.78\pm0.09^{+0.22}_{-0.10}$ for the $\hbp$.
The masses do not differ significantly from the center-of-gravity of
the corresponding $\chi_{bJ}$ states. For the hyperfine splitting we
find $\dmhf=1.62\pm1.52\,\mevm$ for the $\hb$ and
$0.48^{+1.57}_{-1.22}\,\mevm$ for the $\hbp$.

The values of $R$
 comparable with unity indicate that the $\hb$ and
$\hbp$ are produced via an exotic process that violates the
suppression of heavy quark spin-flip. For further  study we
investigate resonant substructure of these decays~\cite{conpa}.
Because of high background Dalitz plot analysis is impossible with
current statistics, therefore we study the one-dimensional
distributions in $M(\hbn\pi)$. We define the $M(\hbn\pip)$ as a
missing mass of the opposite-sign pion, $MM(\pim)$.  We measure the
yield of signal decays as a function of the $MM(\pipm)$ by fitting
the $\mmpp$ spectra in the bins of $MM(\pipm)$. We combine the
$\mmpp$ spectra for the corresponding $MM(\pip)$ and $MM(\pim)$ bins
and we use half of the phase space to avoid double counting.

\begin{figure}[!tbp]
\begin{center}
\includegraphics[width=0.32\textwidth]{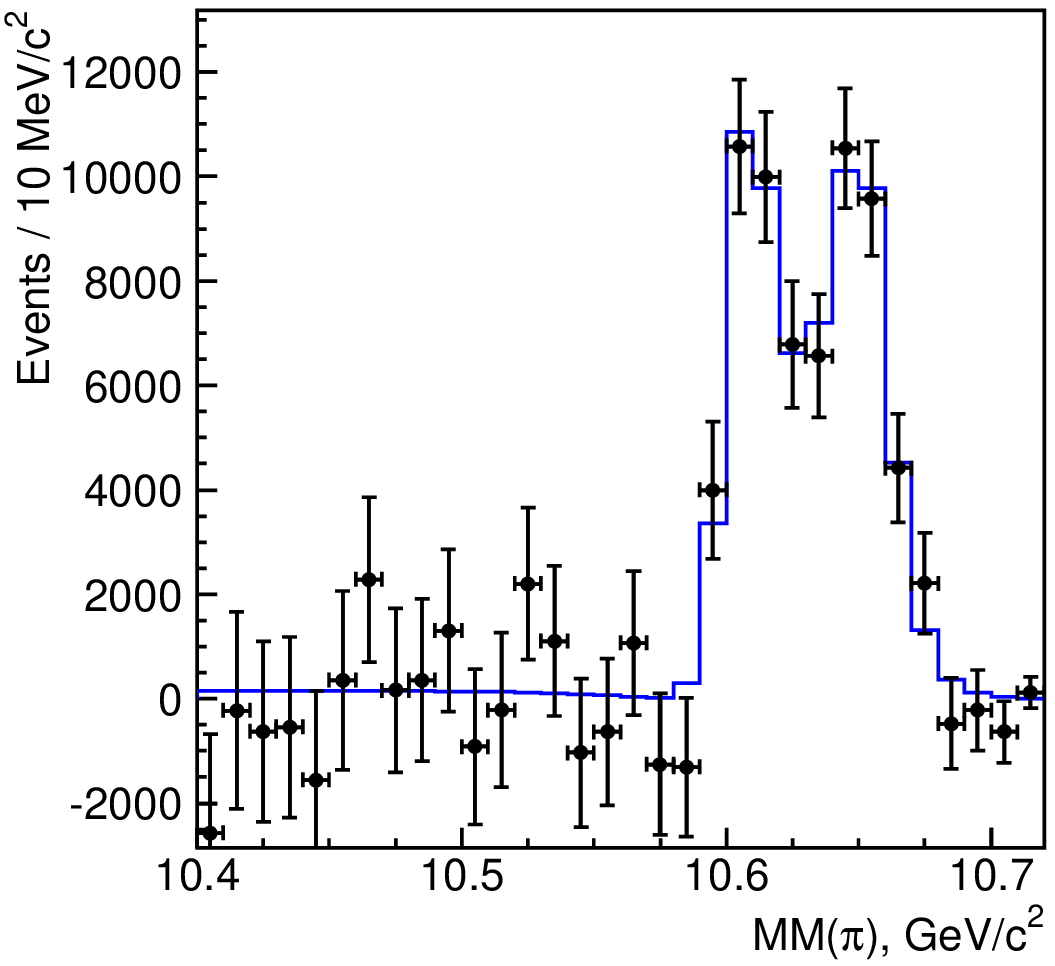}
\includegraphics[width=0.32\textwidth]{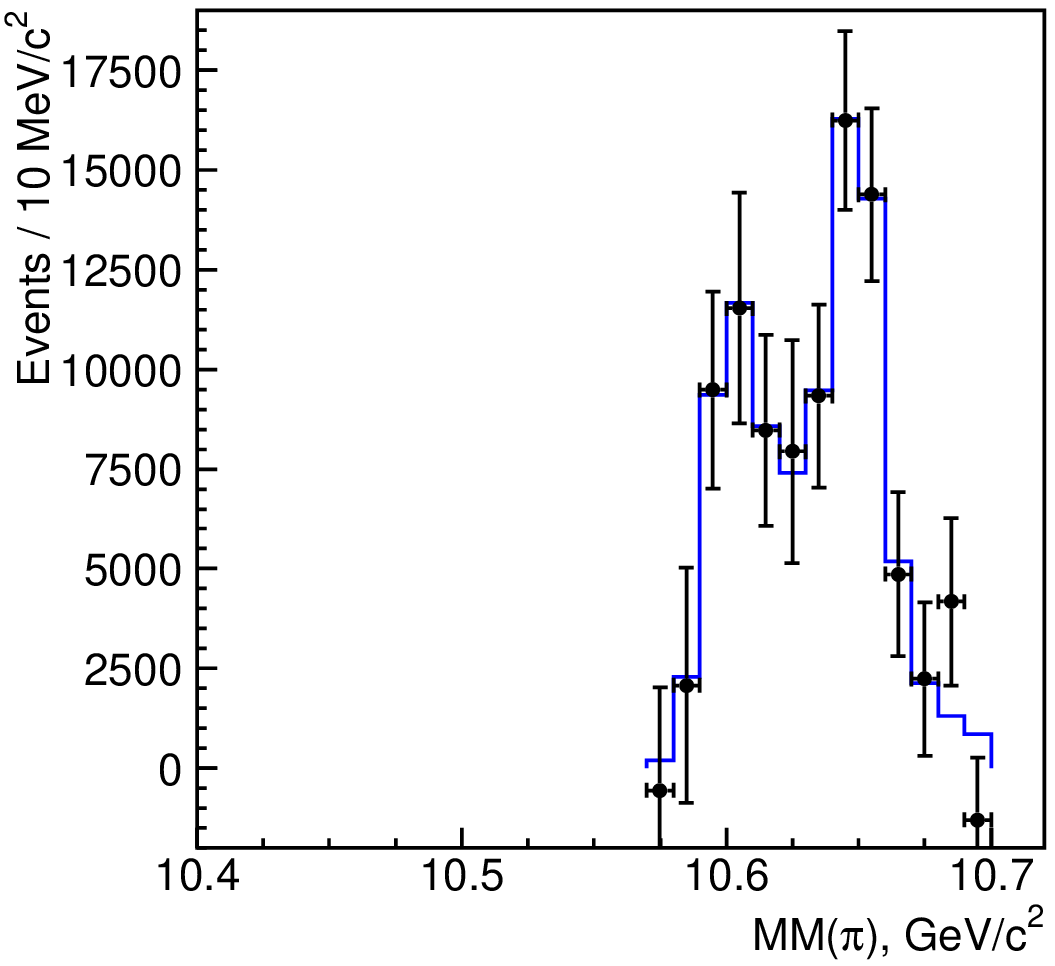}
\caption{Left:  the yield of the $\hb$ as a
  function of $\mmp$ (points with error bars) and results of the fit
  (histogram).  Right: the yield of the $\hbp$ as a function of $\mmp$
  (points with error bars) and results of the fit (histogram).}
\label{nhb_n2s_vs_mmp}
\end{center}
\end{figure}

Results of the fits for the $\hb$  yield as a function of $MM(\pi)$ are
shown in Fig.~\ref{nhb_n2s_vs_mmp}. The $\hb$ yield
 exhibits a clear two-peak structure without
any significant non-resonant contribution. In the following we refer
to these structures as $Z_b(10610)$ and $Z_b(10650)$, respectively.

We perform a $\chi^2$ fit to the $\mmp$ distributions.  We assume
that spin-parity for both $\zbo$ and $\zbt$ is $J^P=1^+$, therefore
in the fit function we use a coherent sum of two $P$-wave
Breit-Wigner amplitudes; we add also a non-resonant contribution.
\begin{equation}
f=A\,|BW(s,M_1,\Gamma_1)+ae^{i\phi}BW(s,M_2,\Gamma_2)
+be^{i\psi}|^2\;\frac{qp}{\sqrt{s}}.
\end{equation}
Here $\sqrt{s}\equiv\mmp$; the variables $A$, $M_k$, $\Gamma_k$
($k=1,2$), $a$, $\phi$, $b$ and $\psi$ are floating in the fit;
$\frac{qp}{\sqrt{s}}$ is a phase-space factor, $p$ ($q$) is the
momentum of the pion originating from the $\Uf$ ($\zb$) decay
measured in the rest frame of the corresponding mother particle. The
results of the fit are shown in Fig.~\ref{nhb_n2s_vs_mmp} and are
summarized in Table~\ref{tab:results}. The non-resonant amplitude is
found to be consistent with zero.
 We find that the hypothesis of two resonances is favored over
the hypothesis of a single resonance (no resonances) at the
$7.4\,\sigma$ ($17.9\,\sigma$) level.  The parameters of the $\zbo$
and $\zbt$ obtained in the fit of $\hb$ and $\hbp$ are consistent with
each other.

\section{Analysis of \boldmath $\Uf \to \Upsilon(1S,2S,3S)\pi^+\pi^-$}

To select $\Uf\to\Un\pp$ candidate events we require the presence of
a pair of muon candidates with an invariant mass in the range of
$8.0~\gevm<M(\mu^+\mu^-)<11.0~\gevm$ and two pion candidates of
opposite charge. These tracks are required to be consistent with
coming from the interaction point. We also require that none of the
four tracks be positively identified as an electron. No additional
requirements are applied at this stage.

Candidate $\Uf\to\Un\pp$ events are identified by the invariant mass
of the $\mu^+\mu^-$ combination and the missing mass
$MM(\pi^+\pi^-)$ associated with the $\pp$ system calculated as $
MM(\pp) = \sqrt{(E_{c.m.}-E_{\pp}^*)^2-p_{\pp}^{*2}}, $ where
$E_{\rm c.m.}$ is the center-of-mass (c.m.) energy and $E^*_{\pp}$
and $p^*_{\pp}$ are the energy and momentum of the $\pp$ system
measured in the c.m.\ frame.

The amplitude analyses of the three-body $\Uf\to\Un\pp$ decays that
are reported here are performed by means of unbinned maximum
likelihood fits to two-dimensional Dalitz distributions.

Before fitting the Dalitz plot for events in the signal region, we
determine the distribution of background events over the Dalitz plot
using events in the $\Upsilon(nS)$ mass sidebands that are refitted
to the nominal mass of the corresponding $\Un$ state to match the
phase space boundaries.

In the sideband Dalitz distributions one can see a strong
concentration of background events in the very low $\pp$ invariant
mass region; these are due to photon conversion on the innermost
parts of the Belle detector. Because of their low energy, these
conversion electrons are poorly identified and pass the electron
veto requirement. We exclude this high background region by applying
the requirements on the $\pp$ invariant mass. For the remainder of
the Dalitz plot the distribution of background events is assumed to
be uniform. The variation of reconstruction efficiency across the
Dalitz plot is determined from MC simulation. The fraction of signal
events in the signal region for each of the three $\Un\pp$ final
states is determined from a fit to the corresponding $MM(\pp)$
spectrum using a Crystal Ball function~\cite{CBF} for the $\Upsilon$
signal and a linear function for the combinatorial background
component.

\begin{figure}[!t]
  \centering
\hspace*{-1mm}
  \includegraphics[width=0.32\textwidth]{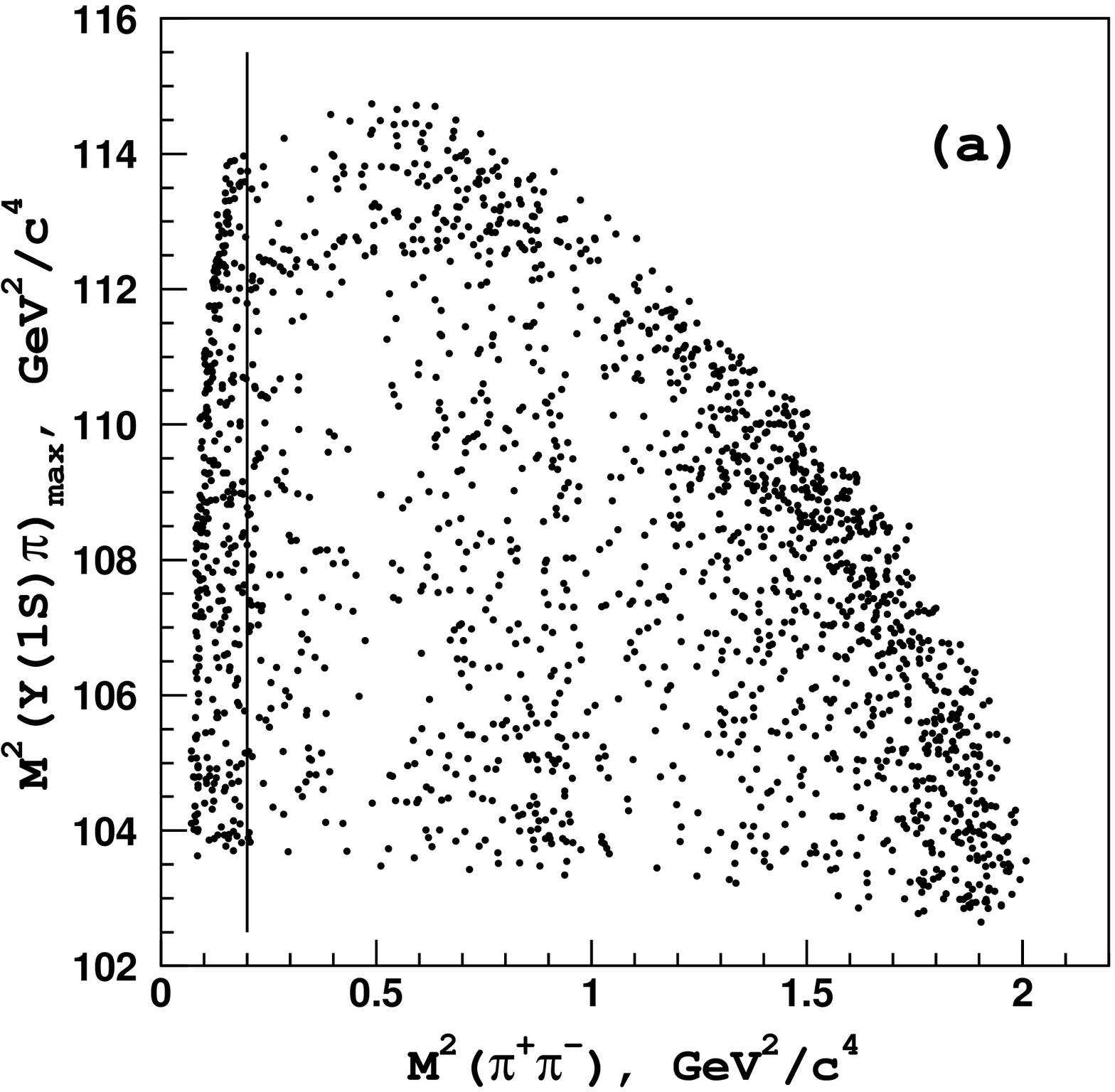} \hfill
  \includegraphics[width=0.32\textwidth]{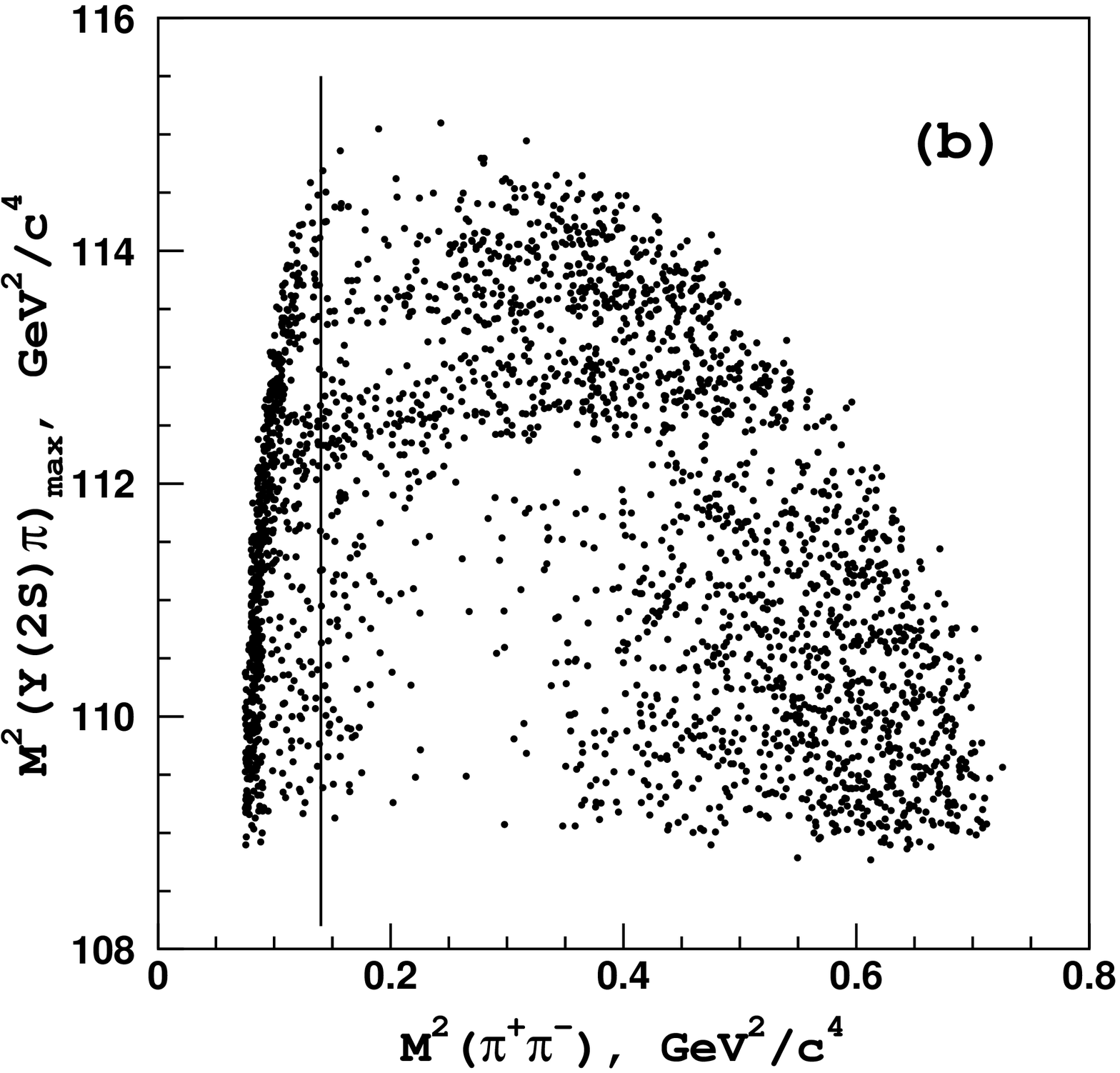} \hfill
  \includegraphics[width=0.32\textwidth]{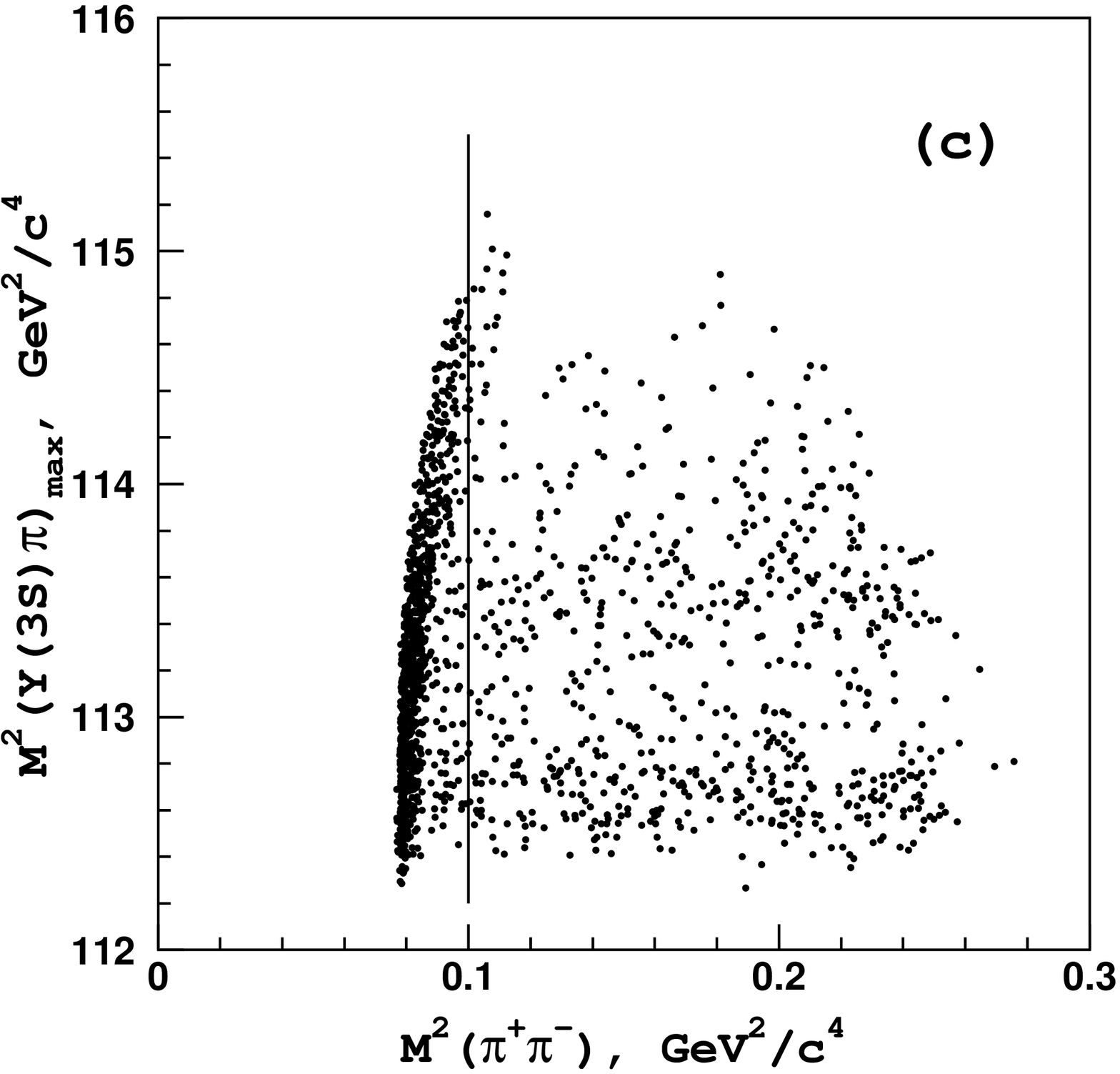}
  \caption{Dalitz plots for $\Un\pp$ events in the (a) $\Uo$; (b)
    $\Ut$; \mbox{(c) $\Uth$} signal regions. Dalitz plot regions to
    the right of the vertical lines are included in the amplitude
    analysis.}
\label{fig:ynspp-s-dp}
\end{figure}

Figure~\ref{fig:ynspp-s-dp} shows Dalitz plots of the events in the
signal regions for the three decay channels under study. In all
cases, two horizontal bands are evident in the $\Un\pi$ system near
$10.61\,\gevm$ ($\sim112.6$~GeV$^2/c^4$) and $10.65\,\gevm$
($\sim113.3$~GeV$^2/c^4$).

\begin{figure}[!t]
  \centering
  \includegraphics[width=0.32\textwidth]{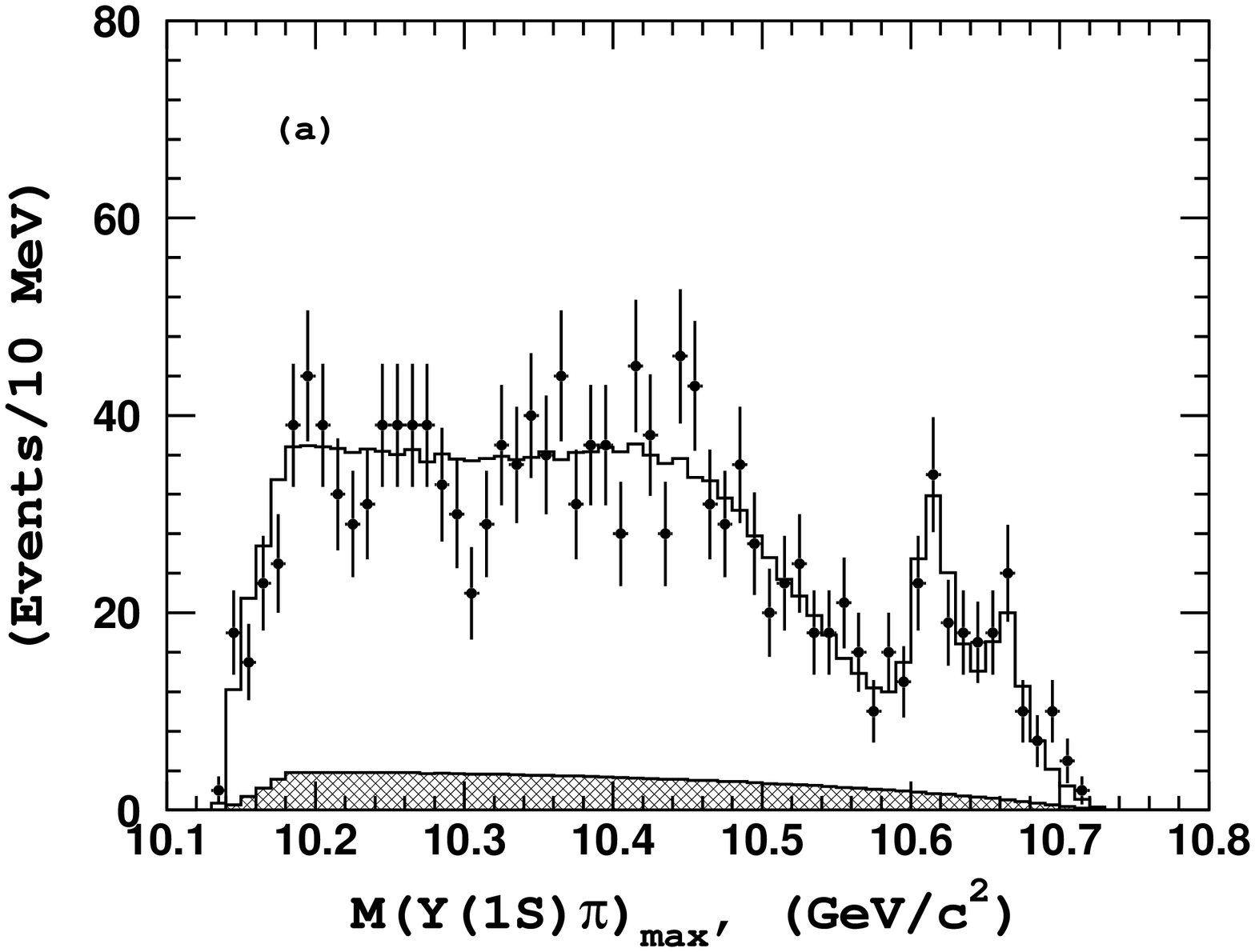} \hfill
  \includegraphics[width=0.32\textwidth]{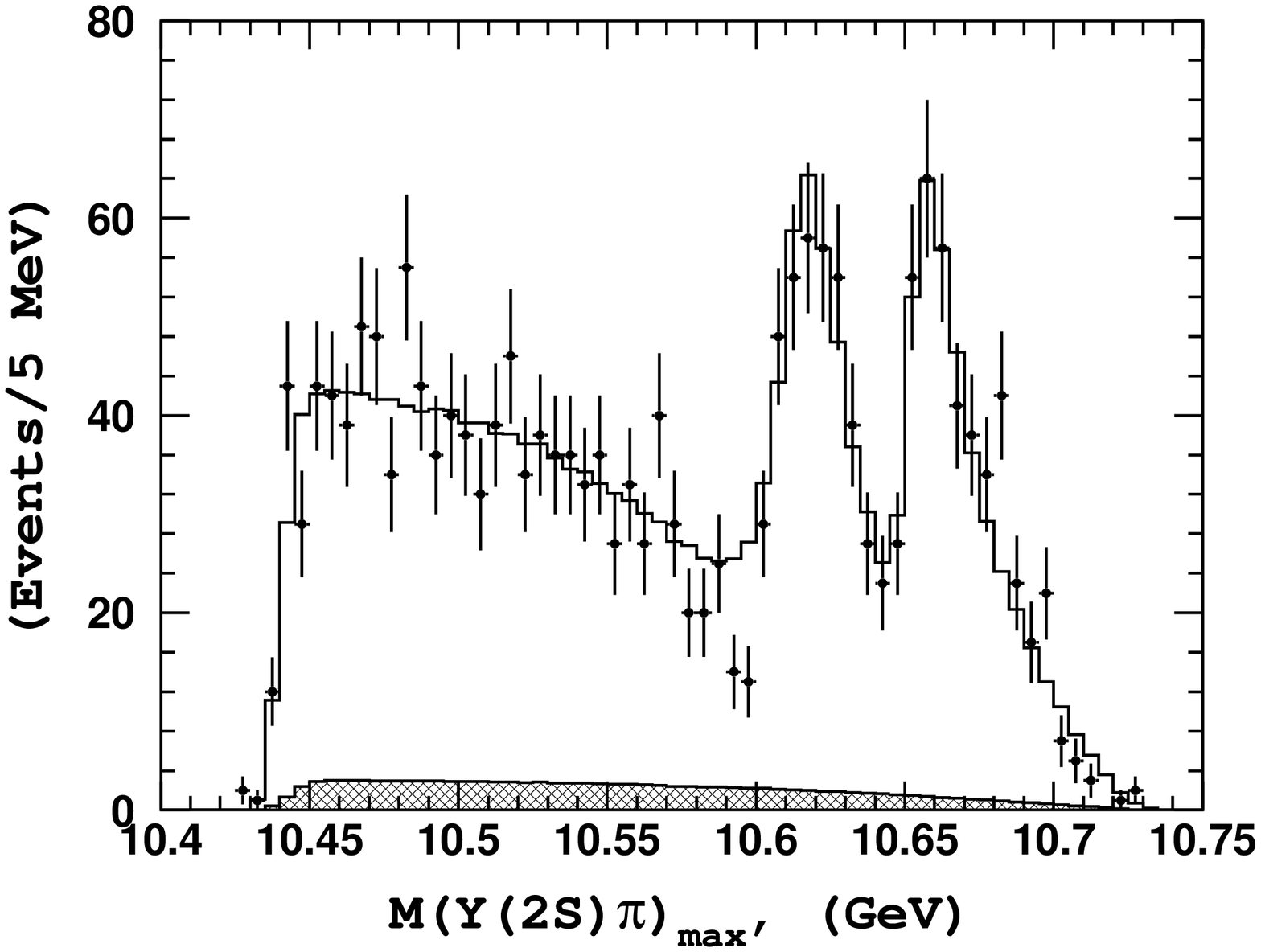} \hfill
  \includegraphics[width=0.32\textwidth]{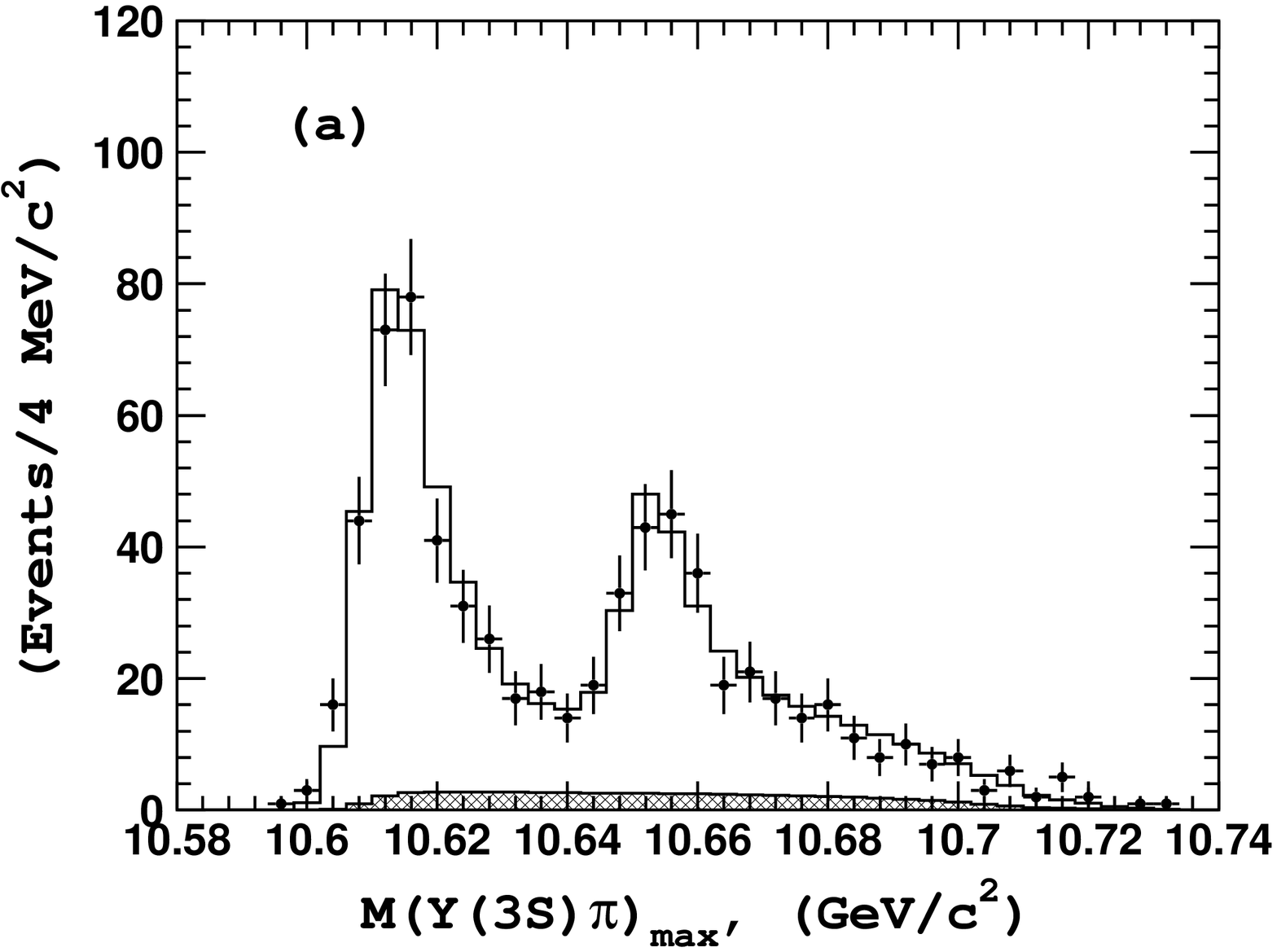} \\
  \caption{Comparison of fit results (open histogram) with
    experimental data (points with error bars) for events in the $\Uo$, $\Ut$ and $\Uth$  signal
    regions.  The hatched histogram shows the background component.}
\label{fig:y3spp-f-hh}
\end{figure}

We use the following parameterization for the $\Uf\to\Un\pp$
three-body decay amplitude:
\begin{equation}
M(s_1,s_2) = A_1(s_1,s_2) + A_2(s_1,s_2) + A_{f_0} + A_{f_2} +
A_{NR}, \label{eq:model}
\end{equation}
where $s_1 = M^2(\Un\pi^+)$, $s_2 = M^2(\Un\pi^-)$. Here we assume
that the dominant contributions come from the amplitudes that
conserve the orientation of the spin of the heavy quarkonium state
and, thus, both pions in the cascade decay $\Uf\to Z_b\pi\to\Un\pp$
are emitted in an $S$-wave with respect to the heavy quarkonium
system. As will be shown later, angular analyses support this
assumption. Consequently, we parameterize the observed $Z_b(10610)$
and $Z_b(10650)$ peaks with an $S$-wave Breit-Wigner function
without $s$ dependence of the resonance width $\Gamma$. To account
for the possibility of $\Uf$ decay to both $Z^+\pi^-$ and
$Z^-\pi^+$, the amplitudes $A_1$ and $A_2$ are symmetrized with
respect to $\pi^+$ and $\pi^-$ transposition. Taking into account
isospin symmetry, the resulting amplitude is written as
\begin{equation}
A_k(s_1,s_2) = a_k e^{i\delta_k} (BW(s_1,M_k,\Gamma_k) +
BW(s_2,M_k,\Gamma_k)),
\end{equation}
where the masses $M_k$ and the widths $\Gamma_k$ ($k = 1,2$) are
free parameters of the fit. Due to the very limited phase space
available in the $\Uf\to\Uth\pp$ decay, there is a significant
overlap between the two processes $\Uf\to Z^+_{b}\pi^-$ and $\Uf\to
Z^-_{b}\pi^+$. We also include amplitudes $A_{f_0}$ and $A_{f_2}$ to
account for possible contributions in the $\pp$ channel from
$f_0(980)$ scalar and $f_2(1270)$ tensor states, respectively.
Inclusion of the $f_0(980)$ state is necessary in order to describe
the prominent structure in the $M(\pp)$ spectrum for the $\Uo\pp$
final state around $M(\pp)=1.0\,\gevm$ (see
Fig.~\ref{fig:y3spp-f-hh}). We also find that the addition of the
$f_2(1270)$ gives a better description of the data at
$M(\pp)>1.0\,\gevm$ and drastically improves the fit likelihood
values. We use a Breit-Wigner function to parameterize the
$f_2(1270)$ and a coupled-channel Breit-Wigner (Flatte)
function~\cite{Flatte} for the $f_0(980)$. The mass and the width of
the $f_2(1270)$ state are fixed at their world average
values~\cite{PDG}; the mass and the coupling constants of the
$f_0(980)$ state are fixed at values defined from the analysis of
$B^+\to K^+\pp$: $M(f_0(980))=950$~MeV/$c^2$, $g_{\pi\pi}=0.23$,
$g_{KK}=0.73$~\cite{kpp}.

Following suggestions given in Refs.\cite{Voloshin:2007dx} and
references therein, the non-resonant amplitude $A_{NR}$ has been
parameterized as
\begin{equation}
A_{\rm NR} = a^{\rm nr}_1\cdot e^{i\delta^{\rm nr}_1} +
             a^{\rm nr}_2\cdot e^{i\delta^{\rm nr}_2} \cdot s_3,
\end{equation}
where $s_3 = M^2(\pp)$ ($s_3$ is not an independent variable and can
be expressed via $s_1$ and $s_2$ but we use it here for clarity),
$a^{\rm nr}_1$, $a^{\rm nr}_2$, $\delta^{\rm nr}_1$ and $\delta^{\rm
nr}_2$ are free parameters of the fit (with an exception of the
$\Uth\pp$ channel as described below).

The logarithmic likelihood function ${\cal{L}}$ is then constructed
as
\begin{equation}
{\cal{L}} = -2\sum{\log(f_{\rm sig}S(s_1,s_2) + (1-f_{\rm
sig})B(s_1,s_2))},
\end{equation}
where $S(s_1,s_2) = |M(s_1,s_2)|^2$ folded with the detector
resolution function (5.6~MeV/$c^2$ for $M(\Un\pi^\pm)$; the $M(\pp)$
resolution is better and is not taken into account since no narrow
resonances are observed in the $\pp$ system), $B(s_1,s_2)=1$ and
$f_{\rm sig}$ is a fraction of signal events in the data sample.
Both $S(s_1,s_2)$ and $B(s_1,s_2)$ are corrected for reconstruction
efficiency.

In the fit to the $\Uo\pp$ sample, the amplitudes and phases of all
of the components are allowed to float. However, in the cases of
$\Ut\pp$ and $\Uth\pp$ the available phase space is significantly
smaller and contributions from the $f_0(980)$ and $f_2(1270)$ are
not well defined.  Thus, in the fit to the $\Ut\pp$ and $\Uth\pp$
signal samples, we fix the amplitudes and relative phases of these
components to the values measured in the fit to the $\Uo\pp$ sample.
Moreover, in the fit to the $\Uth\pp$ sample, we also fix the
$a_2^{\rm nr}$ and $\delta_2^{\rm nr}$ parameters of the $A_{\rm
nr}$ amplitude. Possible effects of these assumptions are considered
while determining the model-dependent uncertainty. Results of the
fits to $\Uf\to\Un\pp$ signal events are shown in
Fig.~\ref{fig:y3spp-f-hh}, where one-dimensional projections of the
data and fits are compared.  To combine $Z^+_b$ and $Z^-_b$ events
we plot $\Un\pi$ mass distributions in terms of $M(\Un\pi)_{\min}$
and $M(\Un\pi)_{\max}$; fits are performed in terms of $M(\Un\pi^+)$
and $M(\Un\pi^-)$. Results of the fits are summarized in
Table~\ref{tab:results}. We try various alternative models to
parameterize the decay amplitude as described in the systematic
uncertainty section. The combined statistical significance of the
two peaks exceeds 10 sigma for all tested models and for all
$\Un\pp$ channels.

\begin{table}[!t]
  \caption{Comparison of results on $Z_b(10610)$ and $Z_b(10650)$ parameters
           obtained from $\Uf\to\Un\pp$ ($n=1,2,3$) and $\Uf\to \hbn\pp$
           ($m=1,2$) analyses. Quoted values are in MeV/$c^2$ for masses, in
           MeV for widths and in degrees for the relative phase. Relative
           amplitude is defined as $a_{Z_b(10650)}/a_{Z_b{10610}}$.}
  \medskip
  \label{tab:results}
\centering { \scriptsize
  \begin{tabular}{lccccc} \hline \hline
 Final state & $\Uo\pp$                   &
               $\Ut\pp$                   &
               $\Uth\pp$                  &
               $\hb\pp$                   &
               $\hbp\pp$
\\ \hline
           $M(Z_b(10610))$ &
           $10609\pm3\pm2$                &
           $10616\pm2^{+3}_{-4}$          &
           $10608\pm2^{+5}_{-2}$          &
           $\mzahb$          &
           $\mzahbp$
 \\
           $\Gamma(Z_b(10610))$ &
           $22.9\pm7.3\pm2$               &
           $21.1\pm4^{+2}_{-3}$           &
           $12.2\pm1.7\pm4$               &
           $\gzahb$   &
           $\gzahbp$
 \\
           $M(Z_b(10650))$ &
           $10660\pm6\pm2$                &
           $10653\pm2\pm2$                &
           $10652\pm2\pm2$                &
           $\mzbhb$      &
           $\mzbhbp$
 \\
           $\Gamma(Z_b(10650))$ &
           $12\pm10\pm3$~                 &
           $16.4\pm3.6^{+4}_{-6}$         &
           $10.9\pm2.6^{+4}_{-2}$         &
           $\gzbhb$           &
           $\gzbhbp$
 \\
           Rel. amplitude                 &
           $0.59\pm0.19^{+0.09}_{-0.03}$  &
           $0.91\pm0.11^{+0.04}_{-0.03}$  &
           $0.73\pm0.10^{+0.15}_{-0.05}$  &
           $\ahb$    &
           $\ahbp$
 \\
           Rel. phase, &
           $53\pm61^{+5}_{-50}$           &
           $-20\pm18^{+14}_{-9}$          &
           $6\pm24^{+23}_{-59}$           &
           $\phihb$         &
           $\phihbp$
\\
\hline \hline
\end{tabular}}
\end{table}

\section{Discussion and Conclusions}

\begin{figure}[!t]
\includegraphics[width=0.80\textwidth]{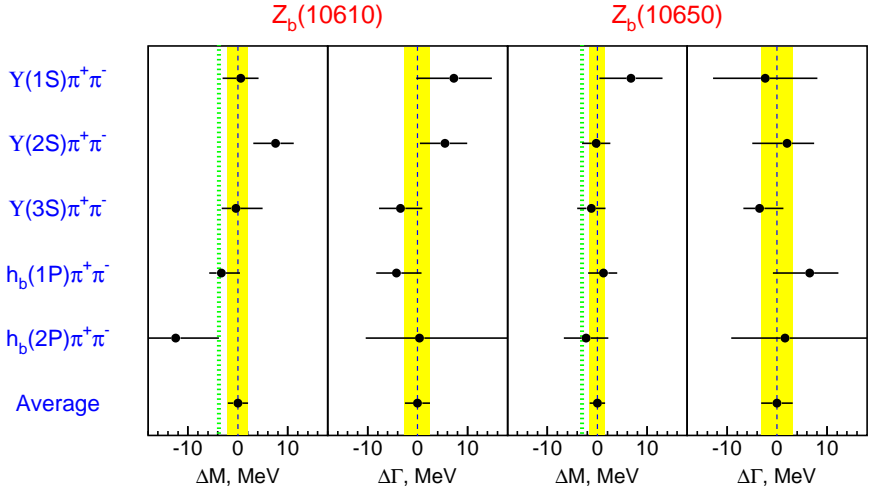}
\caption{Comparison of $Z_b(10610)$ and $Z_b(10650)$ parameters
  obtained from different decay channels. The vertical dotted lines
  indicate $B^*{\overline B}$ and $B^*{\overline B^*}$ thresholds.}
\label{fig:summary}
\end{figure}

In conclusion, we have observed two charged bottomonium-like
resonances, the $\zbo$ and $\zbt$, with signals in five different
decay channels, $\Un\pipm$ ($n=1,2,3$) and $\hbn\pipm$ ($m=1,2$).
Parameters of the resonances as measured in different channels are
summarized in Table~\ref{tab:results}. All channels yield consistent
results as can be seen in Fig.~\ref{fig:summary}.  A simple weighted
averages over all five channels give
$M[Z_b(10610)]=10608.4\pm2.0\,\mevm$,
$\Gamma[Z_b(10610)]=15.6\pm2.5\,\mev$ and
$M[Z_b(10650)]=10653.2\pm1.5\,\mevm$,
$\Gamma[Z_b(10650)]=14.4\pm3.2\,\mev$, where statistical and
systematic errors are added in quadrature.

The measured masses of these states exceed by only a few MeV/$c^2$
the thresholds for the open beauty channels $B^*{\overline B}$
($10604.6$~MeV) and $B^* {\overline B^*}$ ($10650.2$~MeV).  This
``coincidence'' can be explained by a molecular-like type of new
states, {\it i.e.}, their structure is determined by the strong
interaction dynamics of the $B^* {\overline B}$ and $B^*{\overline
  B^*}$ meson pairs~\cite{bondar}.

The widths of both states are similar and are of the order of
$15\,\mevm$.  The $\zbo$ production rate is similar to the $\zbt$
production rate for every decay channel. Their relative phase is
consistent with zero for the final states with the $\Un$ and
consistent with 180 degree for the final states with the $\hbn$.

The $\Uf\to\hbn\pp$ decays seem to be saturated by the $\zbo$ and
$\zbt$ intermediate states; this decay mechanism is responsible for
the high rate of the $\Uf\to\hbn\pp$ process measured recently by
the Belle Collaboration.

Analysis of angular distributions for charged pions~\cite{conpa}
favors the $J^P=1^+$ spin-parity assignment for both $\zbo$ and
$\zbt$. Since the $\Upsilon(5S)$ has negative G-parity, $\zb$ states
will have opposite G-parity due to emission of the pion.

\end{document}